\documentclass[useAMS,usenatbib]{mn2e}
\usepackage{graphicx}

\title[Geological activity at astronomical sites]{Comparative analysis of the impact of geological activity on astronomical sites of the Canary Islands, Hawaii and Chile.}
\author[Eff-Darwich et al.]{A. Eff-Darwich$^{1}$\thanks{E-mail:
adarwich@ull.es}, B. Garc\'{\i}a-Lorenzo$^{2}$, J. A. Rodriguez-Losada$^{1}$, J. de la Nuez$^{1}$
\newauthor
 L. E. Hern\'andez-Guti\'errez$^{3}$, \& M. C. Romero-Ruiz$^{4}$ \\
$^{1}$ Departamento Edafolog\'{\i}a y Geolog\'{\i}a, Universidad de La Laguna\\
C/ Astrof\'{\i}sico Francisco S\'anchez, E-38205, Tenerife, Spain \\
$^{2}$ Instituto de Astrof\'{\i}sica de Canarias, C/Via Lactea S/N, 38305-La Laguna, Tenerife, Spain \\
$^{3}$ Area de Laboratorios y Calidad de la Construcci\'on, Consejer\'\i a de Obras P\'ublicas y Transportes del Gobierno de Canarias\\
C/ Talavera s/n (Llano del Moro), E-38290, Tenerife, Spain \\
$^{4}$ Departamento de Geograf\'{\i}a, Universidad de La Laguna  \\
Campus de Guajara s/n, E-38071, Tenerife, Spain }

\begin{document}


\pagerange{\pageref{firstpage}--\pageref{lastpage}} \pubyear{2002}

\maketitle

\label{firstpage}

\begin{abstract}
An analysis of the impact of seismic and volcanic activity  was carried out at selected astronomical sites, namely the observatories of
El Teide (Tenerife, Canary Islands), Roque de los Muchachos (La Palma, Canary Islands), 
Mauna Kea (Hawaii) and Paranal (Chile) and the candidate 
site of Cerro Ventarrones (Chile). Hazard associated to volcanic activity is low or negligible at all sites, whereas seismic 
hazard is very high in Chile and Hawaii. The lowest geological hazard in both seismic and 
volcanic activity was found at Roque de los Muchachos observatory, in the island of La Palma. 

\end{abstract}


\section{Introduction}

Some of the best astrophysical observatories in the world, namely the Canarian, Chilean and Hawaiian observatories, are located within active geological regions. This is not a coincidence, since topography modelled by tectonic and/or volcanic activity is a main factor controlling the local atmospheric conditions and hence, the sky transparency that defines good astronomical sites. The characterization of astronomical sites is usually based on the analysis of the atmosphere above them \citep{b6}. However, increasingly larger telescopes (10 and 40 metres classes) demand stable structures and buildings, and hence geological activity becomes an important parameter to take into account in astronomical site ranking \citep{b1}. The structures of large telescopes have and will have to withstand the effects associated to seismic and/or volcanic activity, but they also have to minimize the loss of operational time, recalling the extreme precision in the alignment of mechanical and optical components. The west coast of Hawaii experienced in 2006 October 15, a M=6.7 earthquake followed by a M=6.0 aftershock and many smaller aftershocks. There was not significant structural damage at the telescope facilities, however the recovery to full science operability at Subaru, Keck I and II and Gemini North telescopes took several weeks \citep{b12}.

In this work, we present an analysis of seismic and volcanic activity at four world-class observatories, namely Roque de Los Muchachos (La Palma, Canary Islands, 28.75$^\circ$N, 17.89$^\circ$E, 2400 m.a.s.l.), El Teide (Tenerife, Canary Islands, 28.3$^\circ$N, 16.51$^\circ$E, 2380 m.a.s.l.), Paranal (Chile, 24.62$^\circ$S, 70.4$^\circ$E, 2620 m.a.s.l.) and Mauna Kea (Hawaii, 19.82$^\circ$N, 155.47$^\circ$E, 4130 m.a.s.l.) and the candidate site of Cerro Ventarrones (Chile, 24.35$^\circ$S, 70.2$^\circ$E, 2200 m.a.s.l.). We carried out a comparative analysis of the hazard associated to lava flows, volcanic ashfall, seismicity and ground deformation on telescope facilities for a return period of 50 years, assuming this time span corresponds to the approximate lifetime for a large telescope. Flank collapses of island volcanoes \citep{b70}, involving mass movements of tenths to thousands cubic kilometres of rocks, were not considered in this analysis, since their recurrence interval exceedes 100000 years for the Canarian and Hawaiian Islands \citep{b11}. Moreover, the triggering mechanisms of these events and the nature of the mass movements are a 
matter of debate, recalling that humans have not yet witnessed any of these catastrophic events.

\section{Geological context}

\subsection{Tenerife}

Tenerife is the largest island of the Canarian Archipelago and one of the largest volcanic islands in the world. It is located between latitudes 28-29$^\circ$N and longitudes 16-17$^\circ$W, 280 km distant from the African coast. The morphology of Tenerife (see figures 1 and 2) is the result of a complex geological evolution: the  emerged part of the island was originally constructed by fissural eruptions that occurred between 12 and 3.3 Ma \citep{b2}. In the central part of the island, where the observatory is located, the emission of basalts and differentiated volcanics gave rise to a large central volcanic complex, the Las Ca\~nadas Edifice, that culminated in the formation of a large elliptical depression measuring 16x9 km$^2$, known as Las Ca\~nadas Caldera \citep{b3}. In the northern sector of the caldera, the Teide-Pico Viejo complex was constructed as the product of the most recent phase of central volcanism. Teide-Pico Viejo is a large stratovolcano that has grown during the last 175 Ka. The basaltic activity, which overlaps the Las Ca\~nadas Edifice, is mainly found on two ridges (NE and NW), which converge on the central part of the island. Large scale lateral collapses, involving mass movements of hundreds of cubic kilometres of rock, are responsible for the formation of three valleys \citep{b17}, namely La Orotava (0.35-0.65 Ma), G\"uimar (0.75-0.85 Ma) and Icod (0.17 Ma) valleys. Eruptive activity at Mount Iza\~na, where El Teide observatory is located, ceased more than 300 ka, after the formation of La Orotava valley \citep{b16}, whereas recorded eruptive activity in Tenerife has consisted of six strombolian eruptions, namely Siete Fuentes (1704), Fasnia (1705), Arafo (1705), Arenas Negras (1706), Chahorra (1798) and Chinyero (1909). The last three eruptions occurred at the NW ridge, the most active area of the island, together with El Teide-Pico Viejo complex, for the last 50,000 years \citep{b4,b64}.

\begin{figure*}
\includegraphics[width=30pc,angle =0]{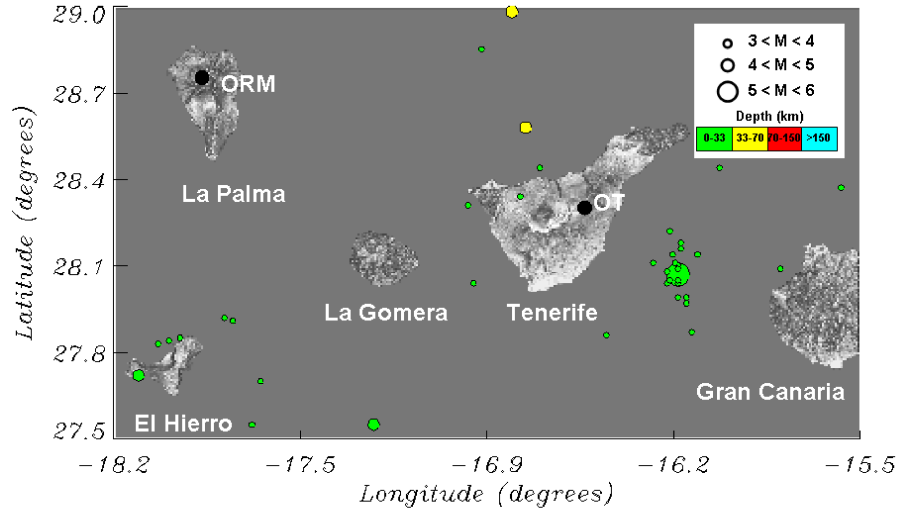}
 \caption{Simplified shaded-relief map of the western Canary Islands. The black filled circles indicate the position of El Teide Observatory (OT) in the island of Tenerife and Roque de los Muchachos observatory (ORM) in the island of La Palma. Filled circles of different sizes and colours indicate the location, magnitude and depth of the earthquakes registered during the period 1973-2008. }
\end{figure*}

\begin{figure*}
\includegraphics[width=40pc,angle =0]{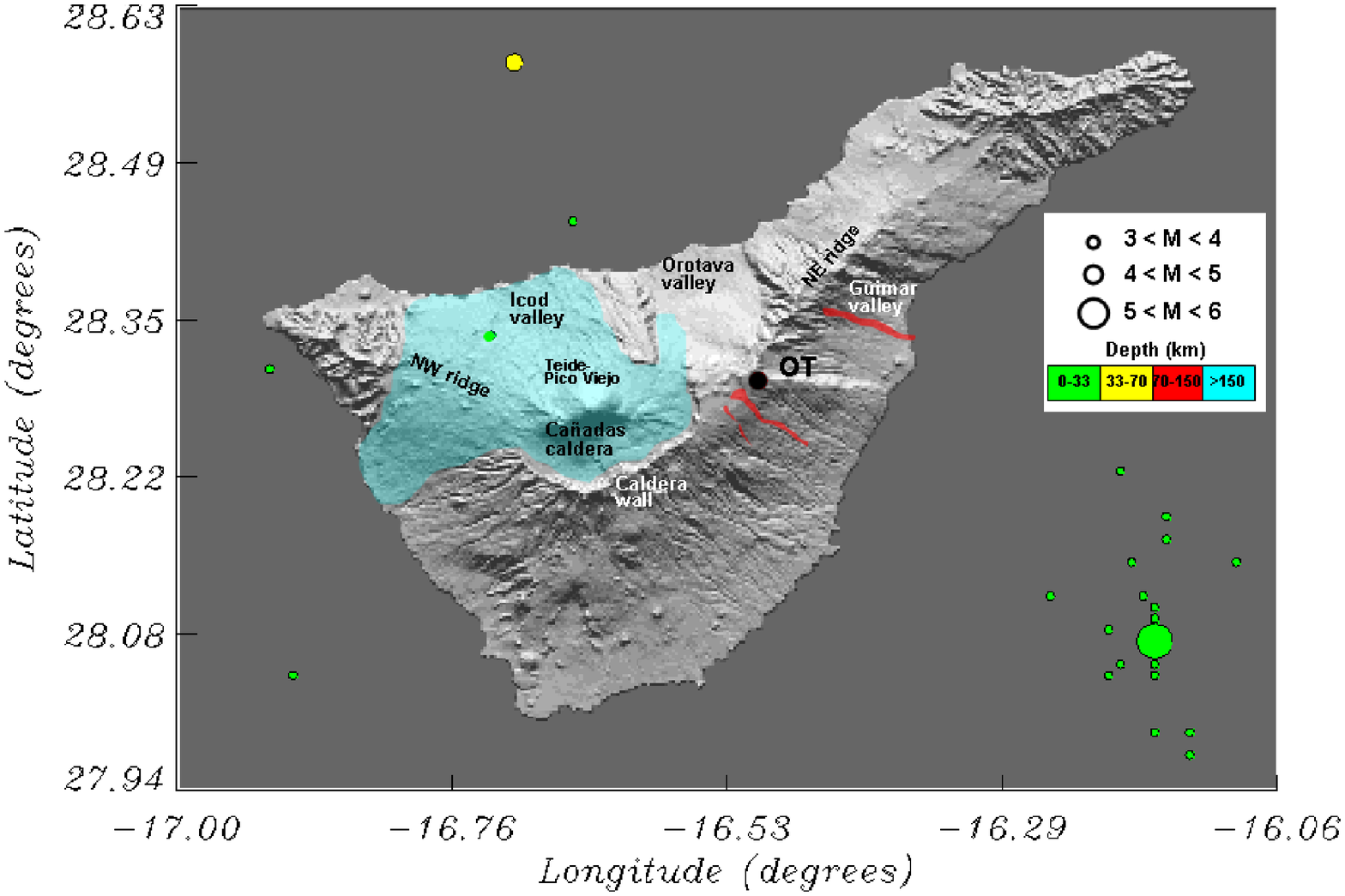}
 \caption{Simplified shaded-relief map of the island of Tenerife, indicating the most important geological features (see text for details). The light blue and red areas indicate the regions affected by recent (last 10 ka) and historical (last 500 years) volcanic activity, respectively. The location of El Teide observatory (OT) is indicated by a black filled circle, whereas filled circles of different sizes and colours indicate the location, magnitude and depth of the earthquakes registered during the period 1973-2008. }
\end{figure*}

\subsection{La Palma}

The island of La Palma is the fifth in extension (706 km$^2$) of the Canary Islands and the second in elevation (2426 m.a.s.l.) after Tenerife (figures 1 and 3). The two main stages of the development of oceanic volcanoes, the submarine and subaerial stages, outcrop in La Palma, since the submarine basement or seamount that was built during the Pliocene (4 to 2 Ma) were uplifted up to 1.3 km above the present sea level \citep{b8,b18}. The subaerial edifice is conformed by a series of overlapping volcanoes \citep{b19,b18,b62,b7}: (1) the Northern Shield of the island includes the Garaf\'\i a shield volcano (1.7 to 1.2 Ma), the Taburiente shield volcano (1.1 to 0.4 Ma) and the Bejenado edifice (0.55 to 0.49 Ma); the Caldera de Taburiente, in the center of the old shield, and Cumbre Nueva ridge formed by a combination of large landslides and erosion; (2) The Cumbre Vieja Volcano (0.4 Ma to present) in the southern half of the island. The Cumbre Vieja ridge is interpreted to be a volcanic rift zone because of the prominent north-south alignment of vents, fissures and faults. Morphologically, Cumbre Vieja is an extension of the Cumbre Nueva ridge and resembles a separate volcano from the Northern Shield as a result of an emigration of the volcanic activity to the south. Historic eruptions on La Palma have lasted between 24 and 84 days and were recorded in 1470/1492 (uncertain), 1585, 1646, 1677-78, 1712, 1949 and 1971 \citep{b63,b20}. The Roque de los Muchachos observatory is located atop the Taburiente Edifice, at the rim of the Caldera de Taburiente, and hence, eruptive activity ceased more than 0.4 Ma ago.

\begin{figure*}
\includegraphics[width=30pc,angle =0]{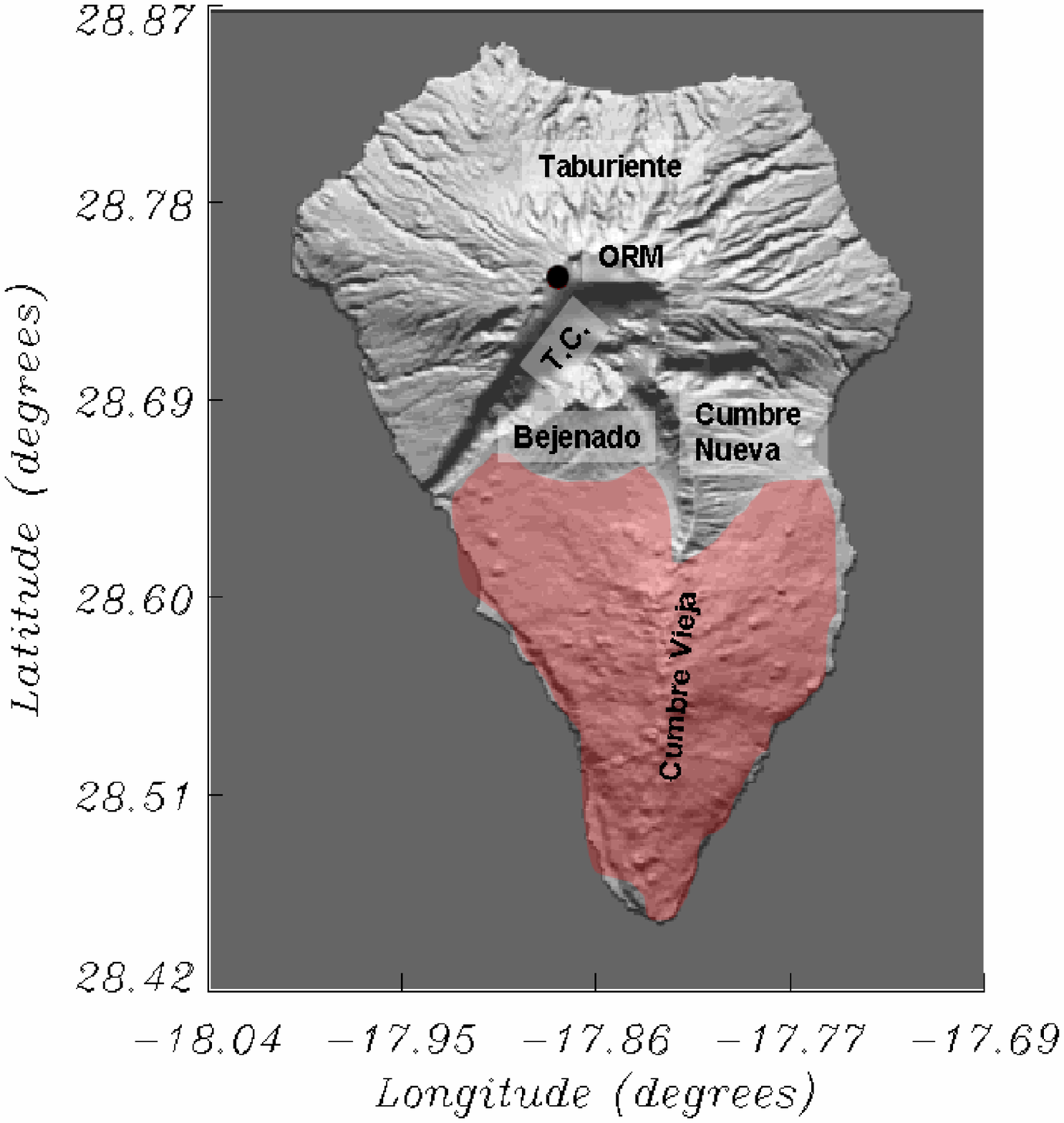}
 \caption{Simplified shaded-relief map of the island of La Palma, indicating the most important geological features (see text for details). The light red area indicates the region affected by recent and historical activity of Cumbre Vieja volcano. The location of Roque de los Muchachos observatory (ORM) is indicated by a black filled circle. There are not earthquakes with magnitudes M$>$3 registered in the area of La Palma during the period 1973-2008.}
\end{figure*}

\subsection{Hawaii}

 The island of Hawaii is the youngest island in a chain of volcanoes that stretches about 5600 kilometres across the northern Pacific Ocean. The island chain results from a magma source that originates deep beneath the crust. The ocean crust and lithosphere above the magma source, within the Pacific tectonic plate, move to the northwest with respect to the deep stationary magma source. Over a span of about 70 million years, new island volcanoes are formed and older volcanoes are carried away from the magma source, erode, and eventually subside beneath sea level \citep{b21}. Hawaiian volcanoes erupt lavas of distinct chemical compositions during four different
stages in their evolution and growth: a) pre-shield stage, representing between 1-2\% of
the total volume, b) shield stage, representing 97-98\% of the total volume, c)
post-shield and post-caldera stage, representing 1\% of the total volume and d)
rejuvenated stage, representing a lot less than 1\% of the total volume.
The island of Hawaii, the largest of the entire Hawaiian chain, consists of five main volcanoes:
Kilauea, Mauna Loa, Mauna Kea, Kohala and Hualalai (Fig. 4). Kilauea and Mauna Loa volcanoes are in the shield stage and erupt
frequently, Hualalai and Mauna Kea volcanoes are in the post-shield stage and erupt every
few hundred to few thousand years, and Kohala is dormant, having passed through the
post-shield stage. Kohala last erupted about 120 ka, Mauna Kea about 3,6 ka, and Hualalai
in 1800-1801 \citep{b22}.

\begin{figure*}
\includegraphics[width=30pc,angle =0]{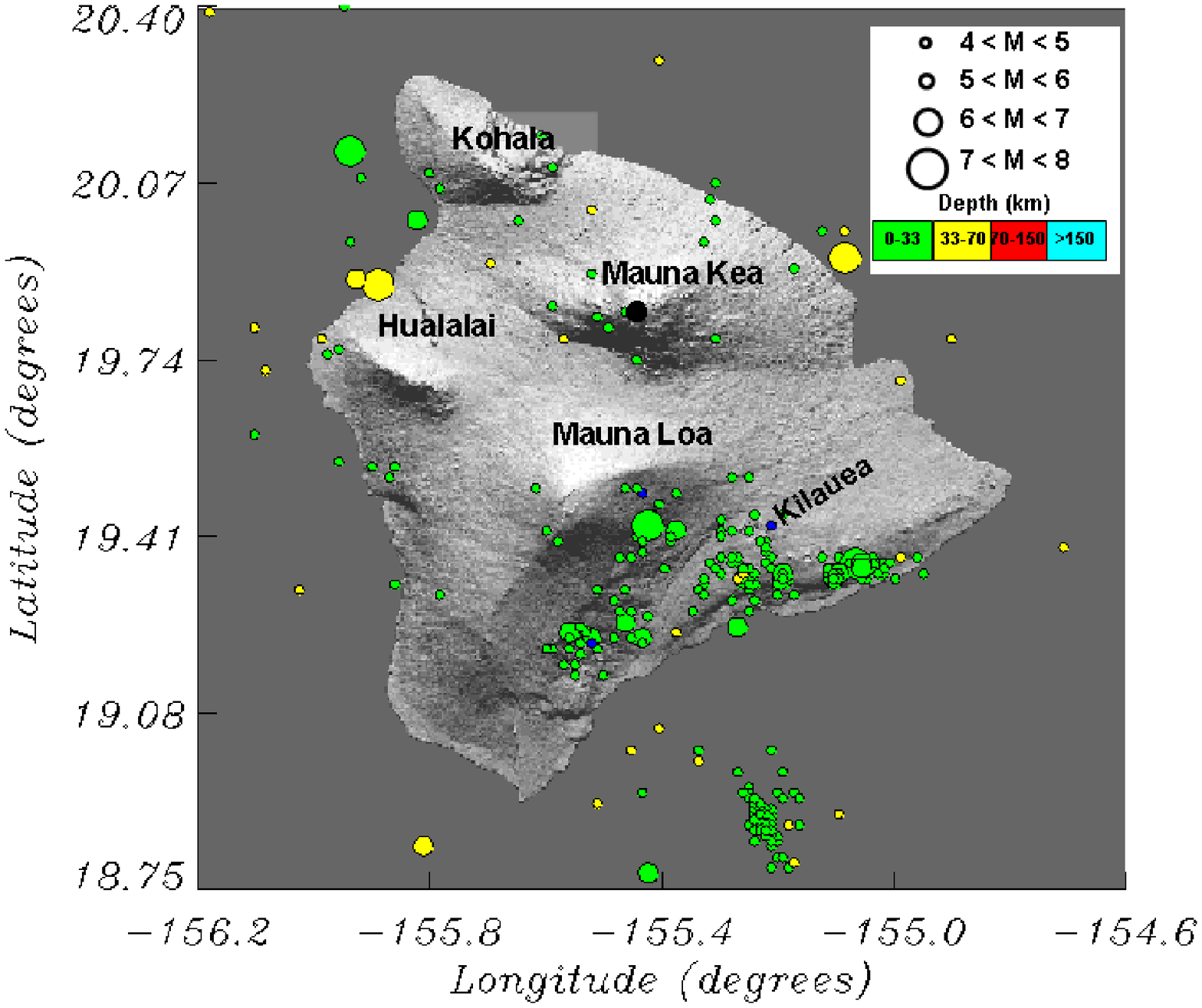}
 \caption{Simplified shaded-relief map of the island of Hawaii, indicating the location of the main volcanoes of the island, namely Kilauea, Mauna Loa, Mauna Kea, Kohala and Hualalai. The location of Mauna Kea observatory is indicated by a black filled circle, whereas filled circles of different sizes and colours indicate the location, magnitude and depth of the earthquakes registered during the period 1973-2008. }
\end{figure*}

\subsection{Central Andes, Chile}

The Andes is a mountain chain which extends along the western edge of the South American plate. In this region, the Nazca Plate thrusts beneath South America at a rate of approximately 64 and 79 mm/year \citep{b23} in an east-north-east direction, forming the Peru-Chile trench in the ocean's floor (Fig. 5). The Andes span a total length of about 9000 km extending from Colombia to the Chile triple junction around 46$^\circ$S, which defines the interaction point of the South American, Antarctic and Nazca plates. The Nazca plate is divided into several segments, whose main difference are convergence direction and dipping angle of the slab \citep{b24}. In general, normal or steep dipping slab segments have angles around 30$^\circ$, whereas in shallow segments this angle does not exceed 15$^\circ$. The angle of the segments also define the presence of volcanic activity, since in shallow segments there have not been volcanic activity at least since the Miocene.

Both Paranal and Cerro Ventarrones are located in the Central Andes region (15$^\circ$S to 27.5$^\circ$S), above a steep dipping segment of the Nazca plate, hence volcanic activity is present, although this takes place more than 200 km eastwards of the sites \citep{b25}. In this sense, volcanic activity ceased in the area of the observatories more than 20 million years ago. The volcanic region of the Central Andes is the largest in the world and it also contains the highest active volcanoes, such as Llullaillaco volcano (6723 m.a.s.l.). Unlike the low viscosity-high temperature magmas of Hawaii, the high viscosity and low temperature of the magmas in the Central andes generate explosive eruptions that eject  large eruptive columns.

\begin{figure*}
\includegraphics[width=40pc,angle =0]{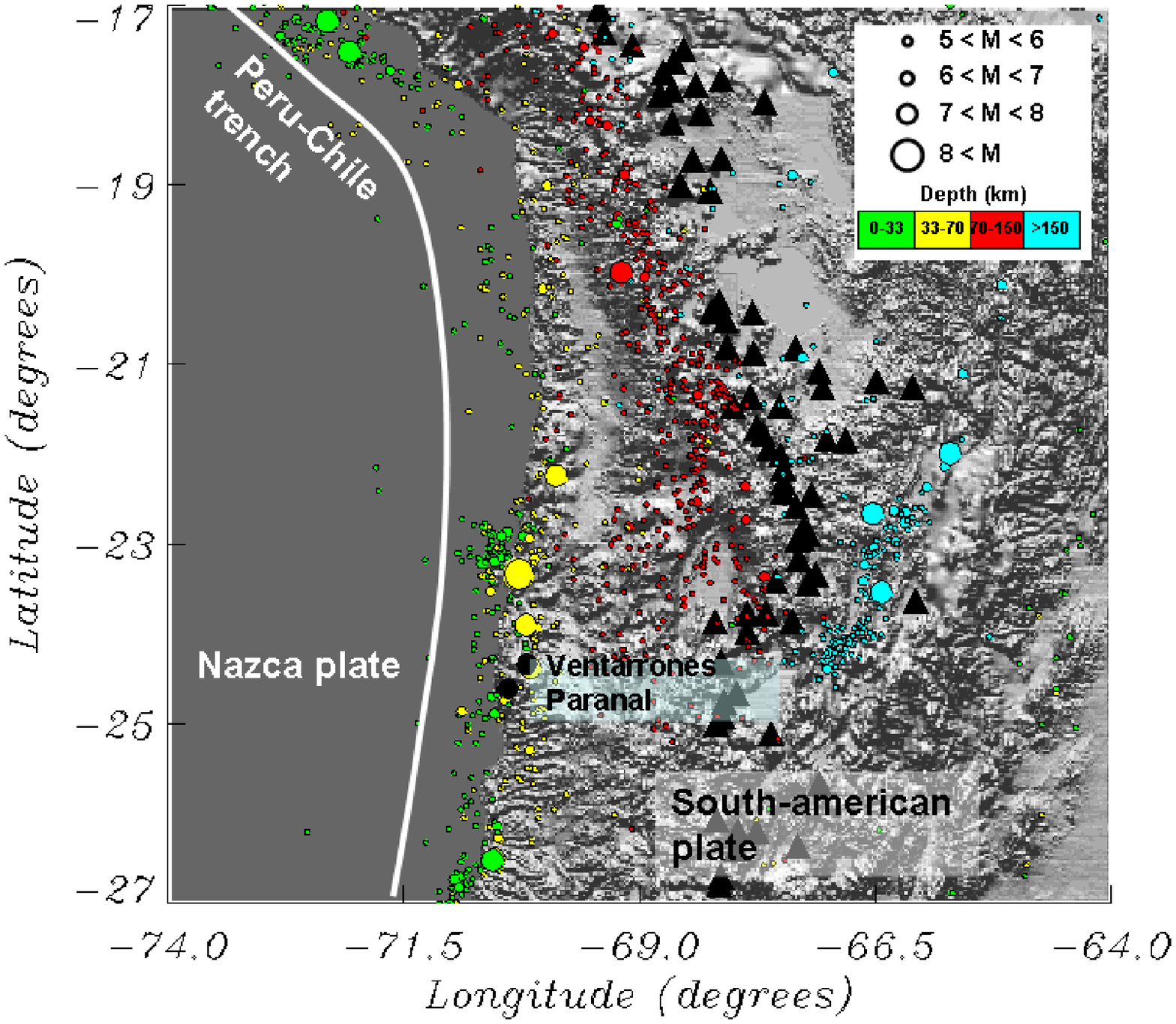}
 \caption{shaded-relief map of the Central Andes, indicating the most important geological features (see text for details), namely the Peru-Chile trench (white solid line) and the volcanic arc (filled black triangles). The location of Paranal observatory and the candidate site of Cerro Ventarrones are indicated by black filled circles, whereas filled circles of different sizes and colours indicate the location, magnitude and depth of the earthquakes registered during the period 1973-2008.}
\end{figure*}

\section{Seismic hazard analysis}

Seismic hazard is definied as the probabilistic measure of ground shaking associated to the recurrence of earthquakes. Seismic hazard maps depict the levels of chosen ground motions that likely will be exceeded in specified exposure times. Hazard asessment for buildings and other structures commonly specifies 10\% chance of exceedance of some ground motion parameter for an exposure time of 50 years, corresponding to a return period of 475 years. The operational and survival conditions of telescope structures will depend on the ground shaking level (seismic hazard) and the seismic design of the structure. Thus, two sites with different levels of seismic hazard will need different seismic designs for the structures to reach the same survival conditions. In this sense, larger seismic hazard means increased building costs \citep{b60}.

 Assessing seismic hazard is the first step in evaluating seismic risk, obtained by convolving seismic hazard with local site effects (topography, geology, amplifications due to soil conditions, ...) and vulnerability factors (buildings and infrastructures, populations, date and time , ...). The basic elements of modern probabilistic seismic hazard are: (1) compilation of uniform seismic databases for historical (pre-1900), early instrumental (1900-1964) and instrumental (1964-present) periods; (2) implementation of a seismic source model to describe the spatial-temporal distribution of earthquakes, using evidence from earthquake catalogues, seismo-tectonics, paleoseismology, geomorphology, active faults, crustal deformation, among others; (3) evaluation of ground shaking as a function of earthquake size and distance; (4) computation of the seismic harzard as the probability of ocurrence of ground shaking in a given time period, to produce maps of seismic hazard. In order to assign a common methodology to infer the probabilistic seismic hazard to the different sites, the data from the Global Seismic Hazard Assessment Program (GSHAP) were analyzed. GSHAP (http://www.seismo.ethz.ch/GSHAP/global/) was launched in 1992 by the International Litosphere Program and implemented in the period 1992-1999 to promote a global homogeneous approach to seismic hazard evaluation \citep{b13}. GSHAP hazard maps depicts Peak-Ground-Acceleration (PGA) with 10 \% chance of exceedance in 50 years, as illustrated in figures 6 to 8 and summarized in Table 1. Although all the sites are located within active volcanic/tectonic regions, seismic activity strongly depends on the geological setting and thus, significant differences in seismicity are expected. This is clearly illustrated in the distribution of epicenters recorded by the National Earthquake Information Center (NEIC), http://neic.usgs.gov, for the period 1973 to 2008 (figures 1 to 5). 

Seismic hazard for the Canary Islands is slightly higher than that reported by the GSHAP in the basis of recent results. In this sense, recent analysis of historical records revealed that earthquakes with intensities of VIII MSK took place during volcanic eruptions in Cumbre Vieja, La Palma and the eruptions of 1704-1705 closer to Mount Iza\~na in Tenerife  \citep{b20}. Moreover, the focal mechanism of the M=5.2 earthquake recorded in 1989 May 9 (the largest earthquake registered in the Canaries) and the aftershock distribution agreeded with the strike of a submarine fault parallel to the eastern coast of Tenerife \citep{b5}. Hence, the activity of this fault and the effect of volcanic activity should be included in the hazard calculation, increasing the PGA from 0.15 m/s$^2$ to 0.56 m/s$^2$ (0.06g) for eastern Tenerife (including El Teide observatory) and 0.5 m/s$^2$ (0.05g) for the rest of the islands. In any case, following the convention of the GSHAP, seismic hazard at the Canary Islands remains at the lowest level. 

The high rate of earthquake occurrence, coupled with the potential for large events, places Hawaii among the areas of highest seismic hazard (figures 4 and 6). Earthquakes are relatively common within the Hawaiian Islands, particularly on the island of Hawaii, as the result of the crustal stresses imparted by the volcanic activity, in particular in Kilauea and Mauna Loa volcanoes. Moreover, the weight of the island has buckled the earth's crust and isostatic adjustements often occur to establish gravitational equilibrium. The earthquakes of 2006 October 15 were caused by isostatic response to the stresses generated in the crust and mantle by the weight of the island. 

The seismic activity near the Chilean sites of Paranal and Cerro Ventarrones (figures 5 and 7) results from the release of stresses generated by the subduction of the oceanic Nazca plate beneath the South American plate. The focal mechanisms of these earthquakes indicate subduction-related thrusting, likely on the interface between these two plates, as illustrated by the relationship between the location of the hypocentres and the distance from the Peru-Chile trench (the longer the distance the deeper the location of the hypocentres). There is a seismic gap (there have not been recorded large earthquakes) between latitudes 18$^\circ$S and 22S$^\circ$S (see Fig. 5) that mark the zones of the 1868 August 14 (M=9) southern Peru and the 1877 May 9 (M=9) northern Chile earthquakes \citep{b26}. Since the estimated recurrence time for large earthquakes ranges between 111 to 264 years \citep{b27}, these regions are likely to experience  a large earthquake in the future.

\begin{figure*}
\includegraphics[width=34pc,angle =0]{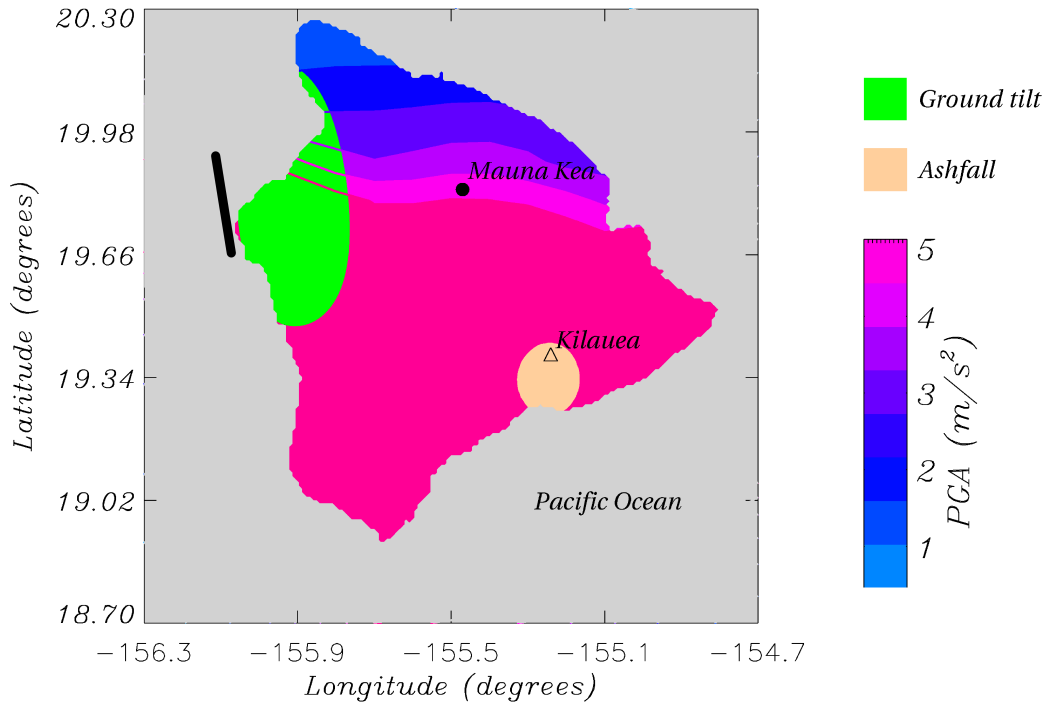}
 \caption{Geological activity map for the island of Hawaii, including Mauna Kea observatory. Green areas indicate the regions where there is at least 1 arcsec of ground tilt induced by the fault dislocation represented by the thick solid black line. Areas that might be covered by at least 0.01 cm of ashfall after an eruption in the Kilauea volcano, are shown in light orange (see text for details).}
\end{figure*}

\begin{figure*}
\includegraphics[width=34pc,angle =0]{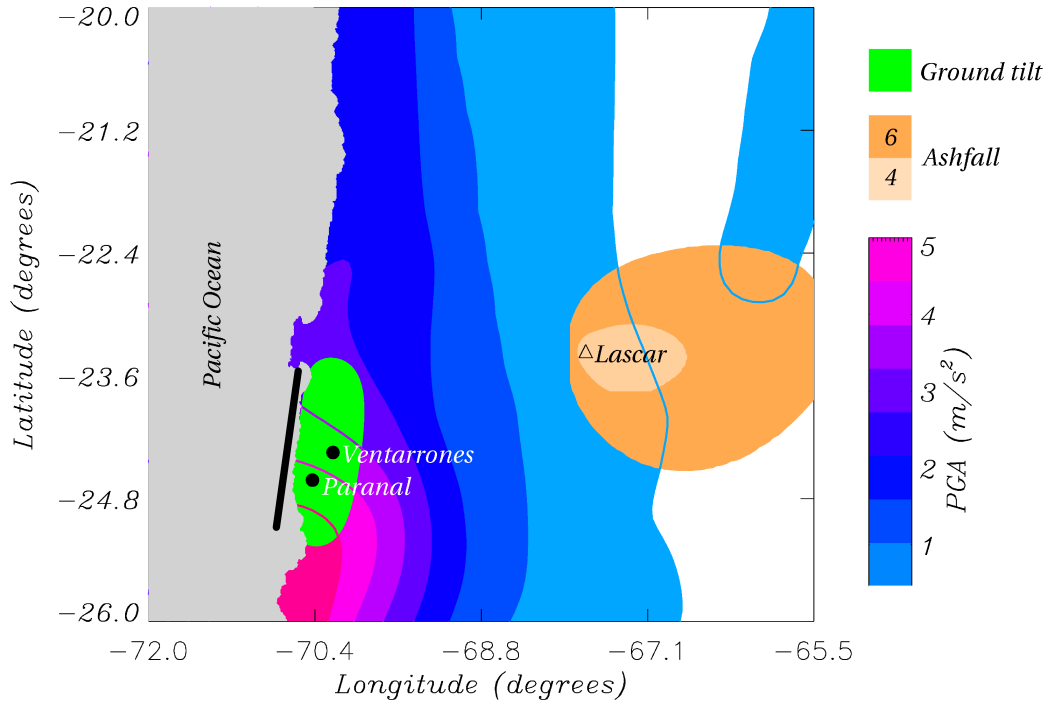}
 \caption{Geological activity map for the Central Andes, including Paranal observatory and the candidate site of Cerro Ventarrones. Green areas indicate the regions where there is at least 1 arcsec of ground tilt induced by the fault dislocation represented by the thick solid black line (see text for details). Areas that might be covered by at least 0.01 cm of ashfall after VEI=4 and VEI=6  eruptions of Lascar volcano are presented in light orange and dark orange, respectively. Peak Ground Accelerations (PGA's) with 10\% chance of exceedance for an exposure time of 50 years are presented in blue-red contours.}
\end{figure*}

\begin{figure*}
\includegraphics[width=34pc,angle =0]{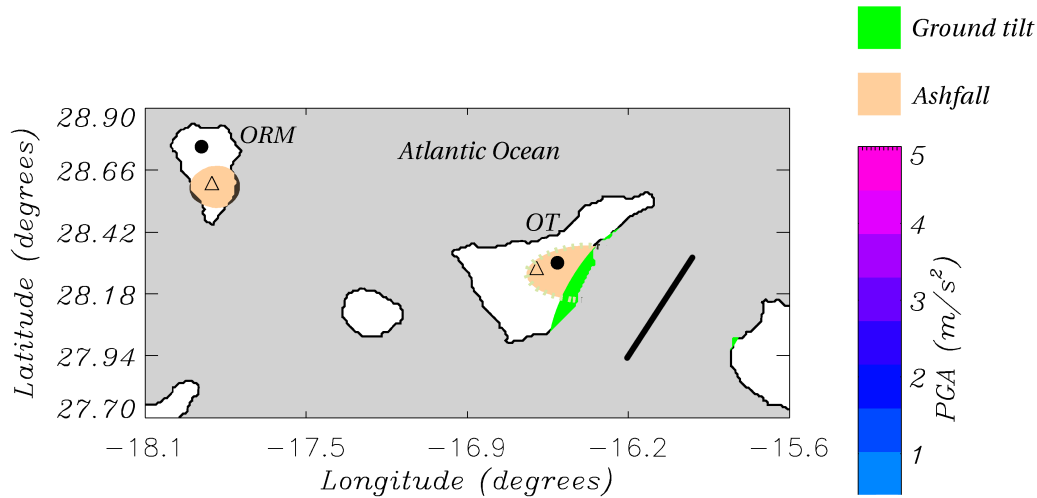}
 \caption{Geological activity map for the western Canary Islands, including El Teide observatory (OT) and Roque de los Muchachos observatory (ORM). Green areas indicate the regions where there is at least 1 arcsec of ground tilt induced by the fault dislocation represented by the thick solid black line. Areas that might be covered by at least 0.01 cm of ashfall after an eruption in the central volcanic complex of Tenerife and Cumbre Vieja volcano in La Palma, are shown in light orange.}
\end{figure*}

\begin{table}
 \centering
  \caption{Seismic hazard expressed in terms of the Peak Ground Acceleration (PGA) with 10\% chance of exceedance for an exposure time of 50 years.}
  \begin{tabular}{@{}ll@{}}
  \hline
  Observatory    &  PGA (g)     \\
 \hline
 Mauna Kea & 0.5 \\
 Paranal & 0.47 \\
 Ventarrones & 0.42 \\
 Roque de Los Muchachos & 0.05 \\
 El Teide & 0.06 \\
\hline
\end{tabular}
\end{table}

\section{Volcanic hazard analysis}

In order to assess the volcanic hazard that might face a site, it is necessary to know the different types of volcanic activity that might be expected and how they present a hazard. Volcanic events may be grouped into two types: on one hand those directly related with eruptions, like lava flows, pyroclastic flows, pyroclastic airfalls and gas emissions; on the other hand those events indirectly related to the eruption like landslides, volcanic 
debris flows (lahars), earthquakes, tsunamis and acid rain. To model the effect of near future eruptions on astronomical sites, only two volcanic phenomena have been considered, namely ashfalls and lava flows, since they might be considered the most likely volcanic hazards affecting all sites, either at short (years) or medium (decades) terms \citep{b28}.

\subsection{Lava flow hazard analysis}

Lava flows are generated by the effusive extrusion of molten rock (magma) that pours onto the Earth's surface. The hazard from lava flows is the damage or destruction by burying, crushing or burning everything in their paths \citep{b15}. The final morphology of the lava flows (length, width, thickness and surace features) is determined by various factors such as the rate of effusion, the slope of the surface onto which is erupted and viscosity and chemical composition of the lava. In general, the higher the content in silica of the lava the more viscuous it becomes. Low silica basaltic lava ($<$52\% SiO$_{2}$) can form fast-moving (15-50 km/h) narrow streams or spread out in broad sheets up to several kilometres wide. In contrast, higher silica lavas (56-70\% SiO$_{2}$) tend to be thick, move slowly and travel short distances. More silicic lavas (65-75\% Si$O_{2}$) tend to form mound-shaped features called domes that rarely exceed 5 km in diameter. Taking into account that lava flows follow valleys and depressions, once potential or actual vents are identified, their likely path can be predicted based on local topography.

In the case of the Chilean sites of Paranal and Cerro Ventarrones, the closest volcanoes of the Central Andes are at least 200 km distant, being significative longer than the typical length of lava flows in the region (30 km). Moreover, volcanic activity at the observatory sites ceased more than 20 million years ago. Hence, hazard due to lava flows is negligible at Paranal and Cerro Ventarrones. 

The Canarian site of Roque de Los Muchachos is emplaced atop the extinct Taburiente volcano, being approximately 1000 metres above and more than 15 kilometres distant from Monta\~na Quemada, the closest eruptive vent of Cumbre Vieja volcano \citep{b63,b20}. In this sense, hazard due to lava flows is also negligible at Roque de los Muchachos observatory. El Teide observatory is emplaced atop Mount Iza\~na, where volcanic activity ceased more than 300 ka. However, the observatory is closer to El Teide-Pico Viejo complex, where there has been an intense volcanic activity in the last 150 ka, whereas several historical effusive basaltic eruptions took place at the beginning of the XVIII century within 10 kilometres from the observatory (Fig. 2).  In any case, the caldera walls protect the observatory from lava flows originated within the caldera area, while local topography and the altitude of the observatory makes the hazard to lava flows from the NE ridge system negligible \citep{b50,b14}. 

There is not lava flow hazard at Mauna Kea observatory due to the eruptive activity of other volcanoes such as Mauna Loa, Kilauea or Hualalai volcanoes, as the result of the topographic protection given by the altitude the observatory is located, above 4100 metres above sea level. However, Mauna Kea is still an active volcano which last erupted 3600 years ago and hence, lava flow hazard is low but larger than in Chile and the Canaries.

Following the convention defined by the United States Geological Survey (USGS), the level of hazard for lava flows has been scaled from 1 to 9, being 1 the most hazardous \citep{b15}. The USGS assigns a hazard level 7 to the area of Mauna Kea observatory, meaning that at least 20 \% of this area has been covered by lavas in the last 10000 years (see Table 2). Hazard level at El Teide observatory was set to 8, since the area covered by lava flows in the last 10000 years is significantly lower than 20 \%, Both the Chilean sites and Roque de los Muchachos observatory in the island of La Palma have a hazard level of 9, since there have not been volcanic activity in the last 60 ka.

\begin{table}
 \centering
  \caption{Level of hazard by lava flows that was calculated following the USGS convention.}
  \begin{tabular}{@{}lll@{}}
  \hline
  Observatory    &  Zone & Last eruption     \\
 \hline
 Mauna Kea & 7 & 3.6 ka  \\
 Paranal and Ventarrones & 9 & $> 1$ Ma  \\
 Roque de Los Muchachos & 9 & 400 ka \\
 Teide & 8 & 300 ka  \\
\hline
\end{tabular}
\end{table}

\subsection{Volcanic ash risk analysis}

Volcanic eruptions commonly eject into the atmosphere fragments of molten lava and rocks (pyroclastics and tephra) that fall back onto the earth surface. An explosive eruption can blast this fragments into the air with tremendous force. The largest fragments (blocks and bombs) falls back to the ground near the vent, usually within 5 km. The smaller rock fragments (lapilli and ash) continue rising into the air, forming a huge eruption column. Eruption columns can be enormous in size and grow rapidly, reaching in some cases more than 20 km in height in less than 30 minutes. They consist of a lower gas-thrust regions and an upper convective region. A column will continue rising convectively until its density equals the surrounding atmosphere. When reaching the troposphere, they will expand laterally and continue growing vertically forming a broad umbrella cloud. Once in the air, the ash and gases of the column are dispersed by the prevailing wind to form an eruption cloud or plume. The strength and direction of the wind and the heigth of the columns are the main factors controlling the transport and final deposition of tephra.

Ash/tephra fall poses the widest-ranging direct hazard from volcanic eruptions. Vast areas (10$^4$ to 10$^5$ km$^2$) have been covered by more than 10 cm of tephra during some large eruptions, whereas fine ash can be carried out over areas of continental size \citep{b29}. Burial by tephra could collapse roofs, break power and communication lines, whereas suspension of fine-grained particles in air affects visibility, could damage unprotected machinery, cause short circuits in electrical facilities and affect comunications.

The spatial extend of tephra fall depends on two factors, namely the wind distribution and the explosivity of the eruption. Wind speeds at different pressure levels for the selected sites were collected from the 
(NCEP/NCAR) Reanalysis database of the National Center for Environmental 
Prediction/National Center for Atmospheric Research. The Reanalysis data span
from  1980 to  2002, being composed of six-hourly and daily U-wind and V-wind 
components. Wind speeds in this database are considered as one of the most reliably 
analysed fields  and these data were validated using radiosonde measurements from the nearest 
station to each site \citep{b40,chueca04}. The vertical distribution of wind direction at all sites (figures 9 to 11) shows a similar behaviour above the 700 mBar pressure level (approximately 2500 m.a.s.l.), being the preferential 
wind direction eastwards. Hence, wind is blowing approximately in the same direction above all the selected sites.

\begin{figure}
\includegraphics[width=18pc,angle =0]{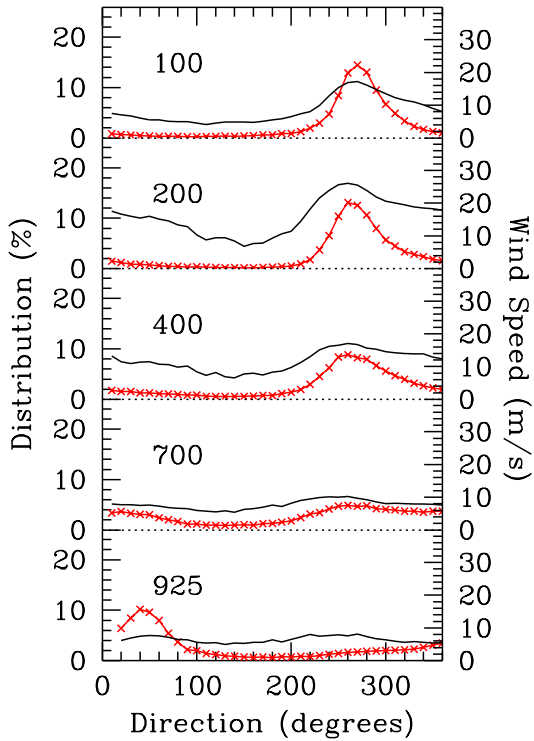}
 \caption{Frequency distribution of wind direction (red lines) and velocity (black lines) on the Canary Islands at five different pressure levels, namely 925, 700, 400, 200 and 100 mbar. }
\end{figure}

\begin{figure}
\includegraphics[width=18pc,angle =0]{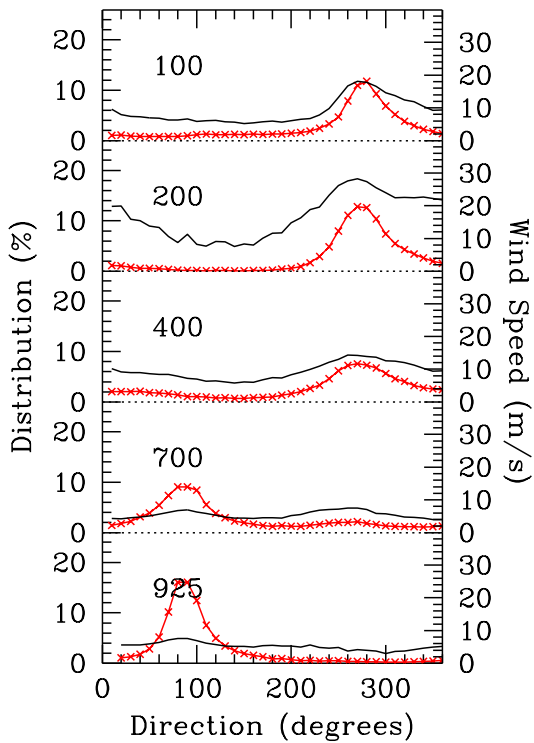}
 \caption{As in Fig. 9, but for the case of the island of Hawaii.}
\end{figure}

\begin{figure}
\includegraphics[width=18pc,angle =0]{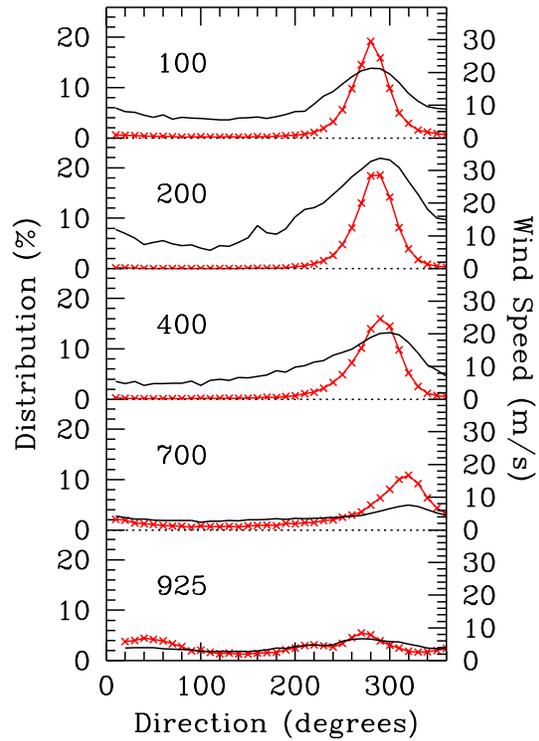}
 \caption{As in Fig. 9, but for the case of Paranal and Cerro Ventarones, Chile.}
\end{figure}

The second major factor affecting the extend of ashfall is the explosivity of the eruption, 
expressed as the Volcanic Explosivity Index (VEI). In this sense, a VEI=3 eruption produces eruptive columns between 3 to 15 km high above the vent, a VEI=4 eruption produces eruptive columns between 10 to 25 km high, whereas eruptions with VEI larger than 4 are associated to eruptive columns higher than 25 km. The Explosivity Index for eruptions in the Central Andes that have been measured in historical times ranges from the VEI=4 sub-plinian eruption of Lascar volcano in 1967 or 1993 \citep{b29} to the VEI=6 eruption of Huaynaputina volcano in February 1600, corresponding to the largest volcanic explosion in South America in historic times.

   The Explosivity Index for eruptions in the island of La Palma is low (VEI$<$3), although eruptive columns up to 4 km high have been recorded in historical times when the uprising magma encountered groundwater, producing phreatomagmatic explosions \citep{b20}. In the case of the island of Tenerife, explosive eruptions might occur in the central part of the island, associated to El Teide-Pico Viejo stratovolcano and its peripherical vents. Last explosive event took place 2 ka and corresponded to the VEI=3 sub-plinian eruption of Monta\~na Blanca volcano \citep{b30}. Due to the complex nature of the central volcanism in Tenerife, other types of explosive eruption have occurred and might occur in the future, namely phreatomagmatic and/or violent strombolian eruptions \citep{b35}. The Explosivity Index for Hawaiian eruptions is similar to that found in the Canarian island of La Palma, namely VEI$<$3, but with chances of phreatomagmatic eruptions like those reported for the Kilauea volcano 2700, 2000, 1100, 600 years ago and the erutptions occurred in 1790 and May 1924 \citep{b31}.

Unlike the hazard analysis for lava flows, there is not established convention to define the hazard due to ash/tephra fall. Hence, a numerical simulator of volcanic asfall, TEPHRA2 \citep{b32}, was used to analyze the extend of ash deposits at each site, considering the typical wind conditions (figures 9 to 11) and the VEI of the volcanoes located in the surroundings of the sites. In this sense, we supposed two different eruptive events (VEI=4 and 6) for the Lascar volcano 
in Chile, a 8 km high eruptive column for a VEI=3 event in Montana Blanca (Tenerife) and 4 km high basaltic eruptive columns for VEI=2 eruptions in Cumbre Vieja (La Palma) and Kilauea (Hawaii). The results of TEPHRA2 (figures 6 to 8) show that El Teide observatory could be affected by a significant deposition of ash (up to 1 cm) due to the prevaling wind conditions, recalling that approximately 60 \% of the time the wind blows eastwards. The Chilean observatory of Paranal and the candidate site of Cerro Ventarrones could be affected by a VEI=6 eruption if the wind blowed westwards, a condition that happens only 0.5 \% of the time. In the case of La Palma and Hawaii, the chances to have significant depositions of ash at the observatory site are negligible, as a result of the distance of the sites relative to the eruptive vents and the direction of the prevailing wind.

In an attempt to assign a probabilistic measure of the hazard associated to ash/tephra fall, we used the methodology implemented by \cite{b33}. This technique assigns the probability that an area might be covered by at least 10 cm of ash during the next 50 years, in terms of the VEI and distance of the volcanic eruption, under the supposition of two different return periods for explosive events, 200 and 2000 years (Table 3). The probabilities of being covered by ash in the astronomical sites of Roque de los Muchachos (La Palma) and Mauna Kea (Hawaii) are negligible, whereas the probabilities in Chile and Tenerife depends on the return period of explosive events. Several explosive eruptions (VEI $>$ 3) have occurred in the Central Andes in the last 100 years \citep{b29}, whereas only two of these events have been reported in Tenerife in the last 35 ka \citep{b64,b35}. The low rate of occurrence of explosive eruptions in Tenerife is also supported by the fact that there are not tephra deposits from the Teide-Pico Viejo complex at the observatory, although this is located in an area with probabilities of being covered by ashfall.

\begin{table}
 \centering
  \caption{Probabilities that an area might be covered by at least 10 cm of ash for an evaluation period of 50 years and two different return periods of eruptive events, namely 200 and 2000 years.}
  \begin{tabular}{@{}lll@{}}
  \hline
  Observatory    &  A &  B    \\
 \hline
Mauna Kea & $0\%$ & $0\%$ \\
Paranal and Ventarrones & $7.5\%$ & $0.75\%$ \\
Roque de Los Muchachos & $0\%$ & $0\%$ \\
El Teide & $12\%$ & $1.2\%$ \\
\hline
\end{tabular}
\end{table}

\section{Ground deformation}

Present-day telescopes reach precisions in pointing and tracking below 1 arcsec, hence stability is required in both the telescope structure and the ground. This is particularly important for long baseline interferometers, where precisions better than 20 $\mu$m are required in the alignment of mechanical and optical components \citep{b36}. \cite{b37} studied the effect of ground tilt at astronomical sites induced by volcanic reactivation of magmatic systems associated to stratovolcanoes and to dike intrusions. These authors concluded  that significant ground tilts of at least 1 arcsec are found within 10 to 15 kilometres from the magmatic system. According to the VEI of the eruptions and the distance to the different astronomical sites, El Teide observatory could be affected by ground tilt larger than 1 arcsec in the case of reactivation of the central volcanic system (El Teide-Pico Viejo stratovolcano and peripherical edifices) and/or a close-by dike intrusion similar to those responsible for the historical eruptions in the island. Moreover, Mauna Kea observatory might also be affected by significant ground tilt as the result of the reactivation of this volcano; however, there is not significant ground tilt induced by activity from Kilauea and/or Mauna Loa volcanoes. Ground tilts at Roque de los Muchachos observatory and the Chilean sites are negligible as the result of the distance to the likely active vents, as observed by \cite{b34} in the Central Andes volcanoes.

We carried out a theoretical analysis to infer the ground deformation, in terms of ground tilt, associated to a dislocation induced by a fault. Following \cite{b39}, we specified nine parameters describing the rectangular fault patch: three centroid coordinates; strike and dip; along-strike length L and down-dip width W; and the slip vector of the left-lateral U1 and up-dip U2 components. These parameters were calculated from the linear relations between earthquake magnitude M and fault morphologies obtained by \cite{b38} for different types of tectonic settings. Following the probabilistic hazard maps calculated by GSHAP, likely earthquakes and the corresponding fault parameters were defined at each site (see Table 4). Absolute vertical displacements and ground tilts associated to the fault dislocations were calculated through the RNGCHN software developed by \cite{b39}. It is observed (figures 6 to 8) that significant ground tilt of at least 1 arcsec might affect the Chilean sites, whereas the effect of ground tilt is less likely in Hawaii and Tenerife and negligible in La Palma.

\begin{table}
 \centering
  \caption{Fault parameters used to calculate ground tilt at the different astronomical sites, namely earthquake magnitude M, Depth of the top of the fault (in km), dip angle of the fault (degrees), length and width of the fault (in km) and dip displacement, U$_{2}$ (in cm). }
  \begin{tabular}{@{}lllllll@{}}
  \hline
  Observatory    &  M &  Depth & Dip & Length & Width  & U$_{2}$   \\
 \hline
Mauna Kea & 7.2  & 30 & 25 & 55 & 20 & 240 \\
Central Chile & 8.0  & 40 & 25 & 160 & 70 & 220 \\
Canary Islands & 6.8 & 25 & 60 & 48 & 10 & 100 \\
\hline
\end{tabular}
\end{table}

\section{Conclusions}

An analysis of geological hazards associated to seismic and volcanic activity (lava flows and ashfall) was carried out at selected astronomical sites, namely El Teide and Roque de los Muchachos in the Canary Islands, Mauna Kea in Hawaii and the Chilean observatory of Paranal and the candidate site of Cerro Ventarrones. A common methodology was used to characterize the geological hazard, expressed in terms of probabilities of occurrence in the next 50 years, recalling that this period of time corresponds to the expected lifetime of a telescope. Large telescopes have to withstand the effects associated to seismic activity, but they also have to minimize the loss of operational time, recalling the extreme precision in the alignment of mechanical and optical components. The hazard from lava flows is the damage or destruction by burying, crushing or burning astronomical facilities  in their paths. Burial by ash/tephra could collapse roofs, break power and communication lines, whereas suspension of fine-grained particles in air affects visibility, could damage unprotected machinery, cause short circuits in electrial facilities and affect comunications. Tectonic and/or volcanic activity might also induce subtle ground tilt that could also result in the disalignment of mechanical and optical components, in particular 
in the case of interferometric observations.

The lowest geological hazard in both seismic and volcanic activity was found at Roque de los Muchachos observatory, in the island of La Palma. Seismic hazard is also low at the other Canarian site, El Teide observatory, since seismic activity in the Canary Islands is low in both number and magnitude of earthquakes. On the contrary, seismic hazard is very high in Paranal and Ventarrones (Chile) and in Mauna Kea (Hawaii). Hazard associated to lava flows during a volcanic eruption is not significant at any site, as the result of low volcanic activity in the regions where the sites are emplaced, topographical protection or distance to the eruptive vents. Hazard associated to volcanic ashfall is negligible at Mauna Kea and Roque de los Muchachos and low to moderate in Tenerife and the Central Andes, depending on the prevailing winds and the still poorly known explosive volcanic activity in these regions. Ground tilt induced by volcanic activity might be significant at Mauna Kea and El Teide observatories, whereas tilt produced by fault dislocation is significant at the Chilean observatories.

\section*{Acknowledgments}

This work has made use of the NCEP Reanalysis data provided by the National Oceanic and 
Atmospheric Administration / Cooperative Institute for Research in Environmental 
Sciences (NOAA / CIRES) Climate Diagnostics Center, Boulder, Colorado, USA, from their 
web site at http://www.cdc.noaa.gov/.

\end{document}